\documentclass[a4paper,12pt]{article}
\usepackage[T2A]{fontenc}
\usepackage[koi8-r]{inputenc}
\usepackage[english,russian]{babel}
\usepackage{indentfirst}
\usepackage{color}
\usepackage{epsfig}
\begin{document}

\begin{center}
{\textbf{Comments to the Dispersion Propertis of the Closed-Ring Accelerating Structure.}}\\
V.V. Paramonov\\
Institute for Nuclear Research of Russian Academy of Sciences, 60-th October 
Anniversary prospect 7a, 117312, Moscow, Russia\\
\end{center}
\bigskip
	\begin{abstract} Closed Ring accelerating Structure (CRS) externally has a set of very attractive properties, 
both in RF parameters and in technological aspects. The structure was proposed for application in compact linear 
accelerators of electrons. Procedure of tuning and structure manufacturing was developed. The structure is now proposed for protons acceleration in medium energy range.\\
For this structure very unusual dispersion properties were declared. It simulates an addutional consideration, 
to understand problem, and additional analysis was performed.\\
Results of analysis indicate particularities of the structure, generating strong hesitations 
for CRS application. 
	\end{abstract}
\bigskip

\begin{center}
{\textbf{Комментарии к дисперсионным характеристикам ускоряющей структуры "замкнутое кольцо".}}\\
В.В. Парамонов\\
Институт ядерных исследований РАН, Проспект 60-летия Октября, 7a, 117312, Москва\\
\bigskip

\end{center}
	\begin{abstract}Кольцевая Бипериодическая Ускоряющая Структура (КБУС), структура 'замкнутое кольцо', 
внешне обладает набором весьма привлекательных характеристик, как электродинамических, так и технологических. 
Цикл работ по исследованию структуры и разработки приемов ее изготовления и настройки  был проведен для ее применения 
в компактном линейном ускорителе электронов.\\ 
Структура предлагается и для применения в линейных ускорителях протонов на участке средних энергий. 
Не типичные дисперсионные свойства структуры стимулируют ее дополнительное изучение.\\
Проведен анализ дисперсионных свойств КБУС, как с помощью метода эквивалентных схем, так и с применением 
прямого численного моделирования частот и распределений полей.  Результаты анализа выявили особенности 
структуры, вызывающие серьезные сомнения в применимости структуры в ускорителе.
	\end{abstract}
\bigskip

\newpage
\tableofcontents
\newpage
\section{Введение} 
\begin{figure}[htb]
\centering
  \includegraphics*[width=150mm]{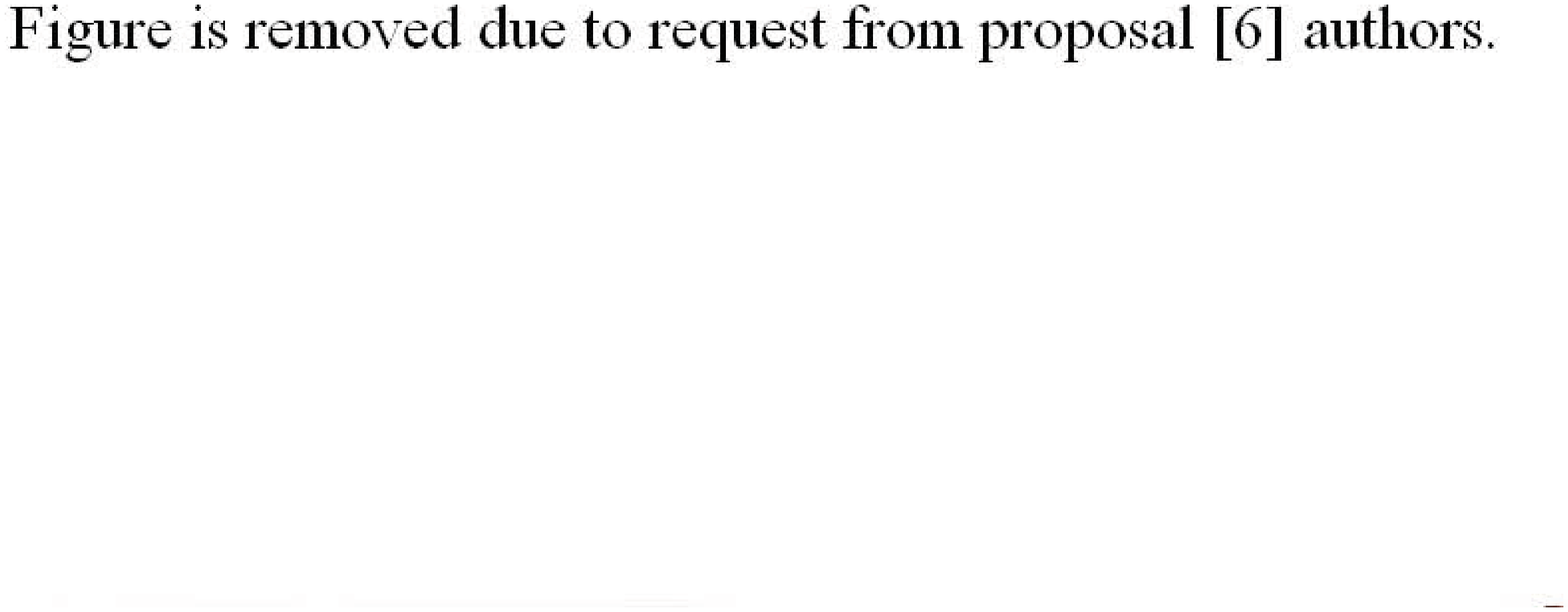}
  \caption{Эскиз секции резонатора N 1 ММФ на базе КБУС, \cite{nano}.}
	\label{kbus_teh}  
\end{figure}

В рамках программы разработки предложения Ускоряющей Структуры (УС) для замены первого резонатора основной части Линейного 
Ускорителя (ЛУ) ионов водорода ИЯИ РАН проводятся исследования по сравнению известных УС и доработке перспективных предложений,
\cite{struct}.\\
\begin{figure}[htb]
\centering
  \includegraphics*[width=90mm]{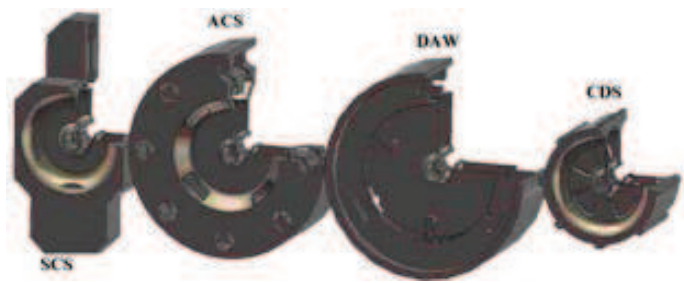}
  \caption{Рассматриваемые УС - SCS, ASC, DAW, CDS.}
	\label{struct}  
\end{figure}
Применение для одиночного резонатора апробированных в ЛУ ионов водорода и имеющих большие поперечные размеры УС, таких как 
УС с боковыми ячейками связи, Side Coupled Structure (SCS), \cite{scs}, УС с кольцевыми ячейками связи, Annular Coupled 
Structure (ACS), \cite{acs}, УС с шайбами и диафрагмами, Disk And Washers (DAW), \cite{daw}, Рис.~\ref{struct}, требует больших затрат 
на изготовление, включая освоение вновь процесса изготовления. \\ 
Рассматривается применение УС Cut Disk Structure, (CDS), Рис.~\ref{struct}, имеющей меньшие поперечные размеры и 
апробированной в ЛУ электронов, \cite{cds}. Адаптация CDS для средних энергий протонов $~ 100~MeV, \beta =0.43,$ требует 
дополнительных исследований, не все получается так хорошо, как хотелось бы, и мы работаем в данном направлении.\\
Для применения в резонаторах ЛУ ИЯИ РАН предложена, \cite{nano},  Кольцевая Бипериодическая УС (КБУС), Рис.~\ref{kbus_teh}.
Предложение выглядит весьма привлекательным технологически, декларируются высокие электродинамические характеристики КБУС. Из предложения 
видны как определенные преимущества КБУС по сравнению с CDS, так и недостатки. Но авторами \cite{nano} декларируются 
весьма необычные дисперсионные свойства КБУС, совместимость которых с основными положениями теории связанных цепей не очевидна. 
Для выяснения дисперсионных свойств резонаторов КБУС и предпринято настоящее рассмотрениее.  
\section{Дисперсионные свойства}
Простейший способ наглядной иллюстрации дисперсионных свойств Кольцевой Цепочки (КЦ) - применение метода эквивалениных схем,
хорошо зарекомендовавшего себя и проверенного для Структур на основе Связанных Резонаторов (ССР) для умеренной величины 
коэффициента связи, \cite{knapp_1}.
\begin{figure}[htb]
\centering
\includegraphics*[width=60mm]{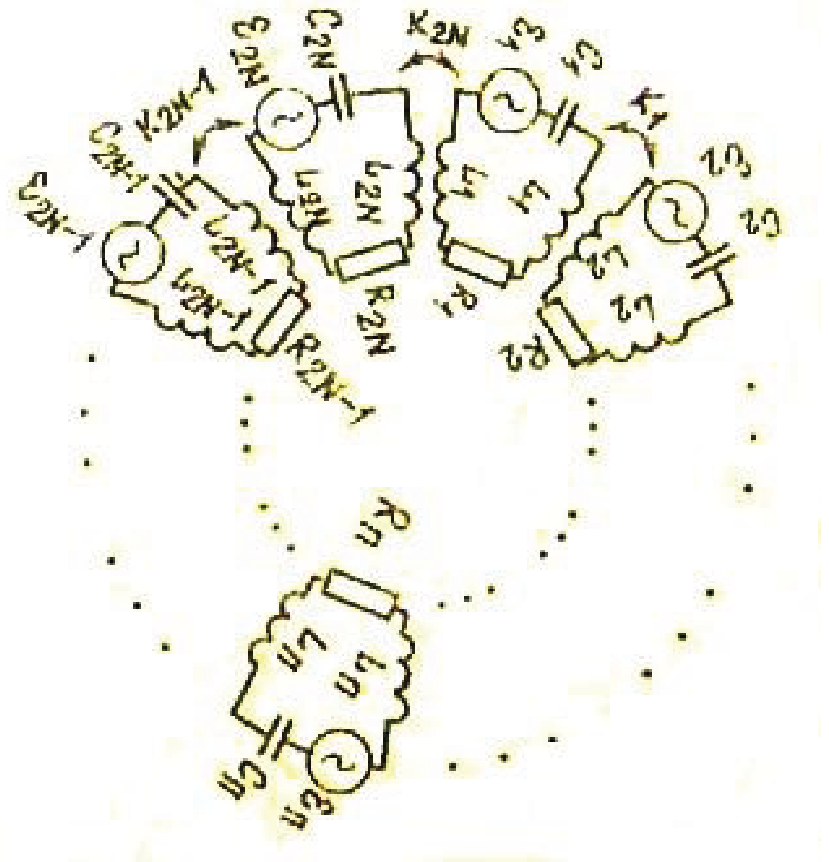}
\caption{Эквивалентная схема замкнутой кольцевой цепочки ячеек.}
\label{eqv_sc}  
\end{figure} 
\subsection{Анализ методом эквивалентных схем}
Эквивалентная схема замкнутой КЦ показана на Рис.~\ref{eqv_sc}. Рассмотрим 
свободные колебания в КЦ предполагая полную идентичность параметров ячеек, частоту всех ячеек $f_c$ и 
коэффициент связи между ними $k$.
\subsection{Периодическая кольцевая цепочка}
Для дальнейшего применения результатов к КБУС, будем иметь ввиду предположение, что КЦ представляет собой последовательность 
их чередующихся ячеек, или резонаторов, двух типов - Ускоряющих Резонаторов (УР) и Резонаторов Связи (РС) и нумерация ячеек 
(резонаторов) $n=1$ начинается с УР. Полная цепочка содержит четное число $2N$ ячеек - $N$ УР и 
$N$ РС. Все ячейки настроены на одну и ту же частоту $\omega_c, f_c=\frac{\omega_c}{2 \pi}$.\\ 
Следуя стандартной процедуре метода эквивалентных схем, получим уравнения для токов $i_n$ в ячейках КЦ:
\begin{eqnarray}
i_1\cdot(1-\frac{\omega_c^2}{\omega^2})+\frac{k}{2}\cdot (i_{2N}+i_{2})=0,\\
\nonumber 
i_2\cdot(1-\frac{\omega_c^2}{\omega^2})+\frac{k}{2}\cdot (i_{1}+i_{3})=0,\\
\nonumber 
..............................................\\
\nonumber 
i_n\cdot(1-\frac{\omega_c^2}{\omega^2})+\frac{k}{2}\cdot (i_{n-1}+i_{n+1})=0,\\
\nonumber 
..............................................\\
\nonumber 
i_{2N-1}\cdot(1-\frac{\omega_c^2}{\omega^2})+\frac{k}{2}\cdot (i_{2N-2}+i_{2N})=0,\\
\nonumber 
i_{2N} \cdot(1-\frac{\omega_c^2}{\omega^2})+\frac{k}{2}\cdot (i_{2N-1}+i_{1})=0,\\
\nonumber 
\label{eq_1}
\end{eqnarray}
где $k$ - коэффициент связи. В общем сучае величина $\omega_c$ включает ВЧ потери как:
\begin{equation}
\frac{\omega_c^2}{\omega^2} = \frac{\omega_c^{'2}}{\omega^2}-\frac{\omega_c^{'}}{\omega} \frac{1}{jQ},
\label{eq_2}
\end{equation}
где $Q$ и $\omega_c^{'}$ - добротность ячейки и собственная частота ячейки без ВЧ потерь, соответственно.
Для наших целей приближение отсутствия ВЧ потерь вполне достаточно.\\
Согласно теореме Флоке, в перодической цепочке решение cуществует в комплексном виде и 
$i_{n+1}=i_n e^{j\theta}$, 
где $0 \leq \theta \leq \pi$ - набег фазы или сдвиг фазы между токами в соседних ячейках.\\
С этим утверждением  о виде решения из центрального 
уравнения системы (1) получаем дисперсионное уравнение и из условия $i_{2N+1}=i_1 \rightarrow 
e^{j(2N+1)\theta_q}=e^{j\theta_q} $ - разрешенные значения $\theta_q$:
\begin{eqnarray}
(1-\frac{\omega_c^2}{\omega^2})+k\cdot cos(\theta)\quad or\quad  \omega=\frac{\omega_c}{\sqrt{1+k cos(\theta})}, \quad f_c=\frac{\omega_c}{2 \pi},\\
\nonumber
\quad \theta_q \cdot 2N =q \cdot 2 \pi, \quad q=0,1,2,...,N, \quad \theta_q = \frac{2 \pi q}{2N}. 
\label{eq_3}
\end{eqnarray}
Набег фазы $\theta_q$ обычно называют видом волны или колебания. Полоса пропускания структуры обычно определеяеся 
как диапазон частот меджу колебаниями $0$ и $\pi$ вида:
\begin{equation}
f_0=\frac{f_c}{\sqrt{1+k}},\quad f_{\pi}=\frac{f_c}{\sqrt{1-k}}, \quad \frac{f^2_{\pi}-f^2_0}{f^2_{\pi}+f^2_0} =k,
\quad f_{\pi}-f_0 \approx k\cdot f_c.
\label{eq_4}
\end{equation} 
В отличие от линейной (не замкнутой цепочки), \cite{knapp_1}, КЦ имеет меньшее, $N+1$, число видов колебаний в полосе 
пропускания. Но система с $2N$ степенями свободы \textbf{должна иметь} $2N$ векторов как решения системы~ (1).
Это фундаментальное условие в теории связанных цепей, см., например, \cite{strelkov}.\\ 
Два решения в виде $e^{j\theta_q n}$ и в виде $e^{-j\theta_q n}$ удовлетворяют системе ~(1) 
и условию замкнутой КЦ, имея одну и ту же частоту. Эти решения представляют собой две Бегущие Волны (БВ), распространяющиеся 
в КЦ в противоположных направлениях - по и против часовой стрелки.\\
Два колебания - две Стоячих Волны (СВ) получаются из двух БВ как:
\begin{equation}
i_{nq}^{SW1} =\frac{e^{j\theta_q n}+e^{-j\theta_q n}}{2} =cos(\frac{2 \pi q n}{2N}), \quad
i_{nq}^{SW2} =\frac{e^{j\theta_q n}-e^{-j\theta_q n}}{2j} =sin(\frac{2 \pi q n}{2N}).
\label{eq_5}
\end{equation} 
Имея одну и ту же частоту, эти колебания имеют различные распределения в пространистве и описывают два независимых 
колебания. Исключение составляют колебания $0$ и $\pi$ видов. Из второго уравнения в ~(5)
для $q=0$ и $q=N$ следует что $i_{nq}^{SW2}=0$, отражая хорошо известное утверждение - для $\theta=0$ и $\theta = \pi$ 
прямая БВ идентична обратной БВ и обе они идентичны колебанию СВ.\\
Заключение о двух независимых распределениях для каждого внутреннего, $0 < \theta_q < \pi$, вида колебаний можно 
продемонстрировать более наглядно.
Нумерация ячеек в цепочке Рис.~\ref{eqv_sc} $n=1$ начинается с УР и пусть существует только одно решение,
например $i_{nq}^{SW2} =sin(\frac{2 \pi q n}{2N})$. Повернем КЦ на угол $\frac{2 \pi}{2N}$. Тогда КЦ переходит сама в себя 
и ничего, кроме нумерации, не изменяется. Частоты колебаний, описываемые системой ~(1), остаются прежними. 
И решение, но с нумерацией, начинающейся тепеть с РС $n^{'}=1$, обязано существовать, 
$i_{n^{'} q}^{SW2} =sin(\frac{2 \pi q n^{'}}{2N})$. Но это \textbf{новое} решение, существующее на той же частоте, отличается от 
начального, оно описывает \textbf{другое} распределние поля.\\ 
В замкнутой КЦ для каждого внутреннего вида на одной и той же частоте существуют \textbf{два} 
независимых колебания. Это означает, что каждый внутренний вид колебаний в замкнутой КЦ \textbf{дважды вырожден} по частоте.
Таким образом устраняется противоречие и система с $2N$ степенями свободы имеет $2N$ незавиcимых решений.\\
Для замкнутой КЦ этот вывод известен. Замкнутую КЦ представляют собой резонаторы магнетронов, 
свойства которых очень хорошо исследованы для широких практических применений. Подтвержение сделанным в данном 
подразделе выводам можно получить, например, в \cite{mag}, страница 170. \\  
\subsection{Нарушения периодичности КЦ применением различных ячеек связи.}
Обычно ССР работают на $\theta =\frac{\pi}{2}$ виде колебаний. В этом случае в соседних УР реализуется 
противопоположные направления ускоряющего поля. В идеальной замкнутой КЦ при нечетном $N$ колебание вида $\theta =\frac{\pi}{2}$ 
существовать \textbf{не может}. Это легко видеть из последнего уравнения системы ~(1). Для $\theta =\frac{\pi}{2}$, 
$i_{2N-1}=i_1=1$ и это уравнение нельзя удовлетворить невозбужденной РС с $i_{2N}=0$.\\
Как видно из ~(3), колебание $\theta =\frac{\pi}{2}$ имеет частоту $f_{\frac{\pi}{2}} = f_c$.
В замкнутой КЦ с нечетным $N$ \textbf{можно} организовать колебание с частотой $f_c$ и распределение подобным  $\theta =\frac{\pi}{2}$ виду 
нарушением периодичности цепочки и введением одного Инвертирующего РС (ИРС).
\subsubsection{Прямые и Инвертирующие Резонаторы (ячейки) Связи} 
\begin{figure}[htb]
\centering
  \includegraphics*[width=130mm]{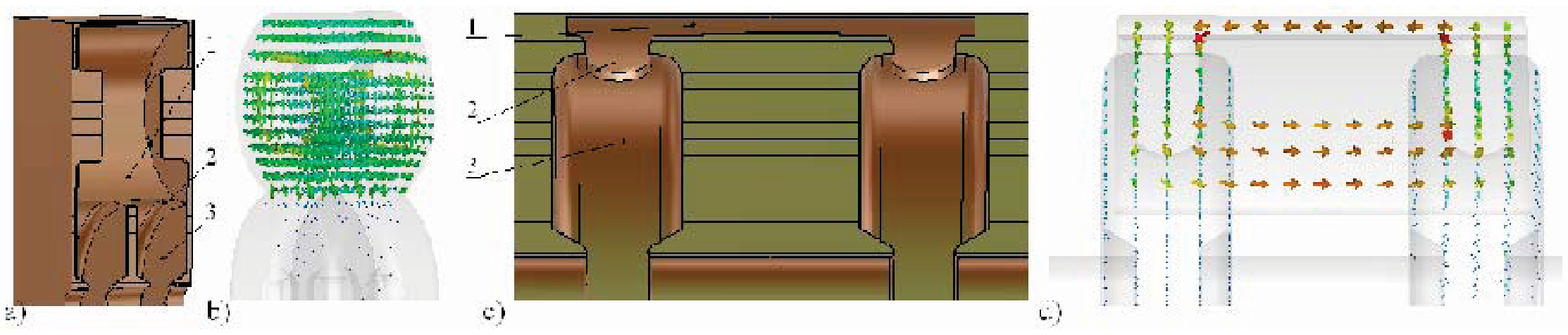}
  \caption{Прямая ячейка связи (a) с распределением поля (b) в сравнении с инвертирующей ячейкой связи (c) с распределением 
поля (d). 1 - ячейка связи, 2 - щель связи, 3 - ускоряющая ячейка.}
	\label{dcc_rcc}  
\end{figure}

При выводе системы (1) предполагалось, что коэффициенты связи по обе стороны РС одинаковы. Для иллюстрации различия при 
применении в методе эквивалентных схем, рассмотрим два типа РС, показанных на Рис.~\ref{dcc_rcc}.\\ 
На Рис.~\ref{dcc_rcc}a показана боковая ячейка связи (РС) структуры SCS, \cite{scs}. В этом РС возбуждается колебание 
$TM_{010}$ и магнитное поле на соседних щелях связи направлено одинаково, Рис.~\ref{dcc_rcc}b. В уравнениях 
для эквивалентной схемы мы должны ставить одинаковый знак коэффициента связи на обеих сторонах РС. В дальнейшем в тексте 
будем называть такие ячейки-резонаторы Прямыми РС - ПРС.\\ 
Секторный призматический РС КБУС, \cite{zav1}, на Рис.~\ref{dcc_rcc}c возбуждается на колебании $TE_{101}$ и магнитные 
поля на щелях связи имеет противоположное направление, Рис.~\ref{dcc_rcc}d. Для правильного применения в эквивалентной 
схеме мы должны применять разные знаки коэффициента связи на противоположных сторонах РС. В дальнейшем в тексте 
будем называть такие ячейки-резонаторы Инвертирующим РС - ИРС.\\ 
Применяя только один ИРС как $(2N-1)$-й РС, мы получаем в замкнутой КЦ колебаний с частотой $f_{\frac{\pi}{2}} = f_c$ 
и противоположными направлениями ускоряющего поля в соседних РУ. Последнее уравнение в (1) теперь удовлетворяется. Вырождение 
колебания $\frac{\pi}{2}$ сохраняется, возможны решения как $cos(\frac{2 \pi n}{2})$ так и $sin(\frac{2 \pi n}{2})$.
Но теперь существование колебания $\pi$ вида запрещено последним уравнением в (1).\\
Введение даже одного ИРС в замкнутую КЦ является нарушением периодичности, выражения (3) для дисперсионной кривой и (5) 
для распределений полей могут быть использованы только как предварительные оценки. Параметры полной замкнутой КЦ теперь 
могут быть рассчитаны только в прямом численном моделировании.  
\subsubsection{Примеры численного моделирования для эквивалентной схемы КЦ}
Система уравнений (1) формирует обобщенную задачу на собственные значения: 
\begin{equation}
A \cdot I - \omega^2 \cdot B \cdot I =0,
\label{eq_6}
\end{equation}
где $I= (i_1, i_2, ..., i_{2N-1}, i_{2N})$ - вектор решения, токи в ячейках КЦ, $A$ - диагональная матрица  
$a_{ll}=\omega_0^2$ и $B$ - матрица жесткости, отражающая связь прилегающих ячеек в (1) и $b_{ll}=1$.\\
Используя стандартные программы, например подпрограмму F02BJF библиотеки алгоритмов Fortran 77 NAG, \cite{nag}, 
мы можем непосредственно найти решение системы (1). В этом случае мы не используем никаких предположений 
и напрямую решает систему (1), которая описывает распределения полей и частот в КЦ. Таким образом можно подтвердить, 
или не подтвердить результаты и выводы, полученные аналитически в предыдущем подразделе.\\ 
Результаты численного рассчета представлены в таблицах. Чтобы иметь длину таблиц в разумных пределах, рассмотрим 
два варианта с умеренное числом ячеек - четное $N=8$ и нечетное $N=9$ число РУ в замкнутой КЦ.\\
Предполагается что частота ячеек КЦ $f_c= 991.0 МГц$ совпадает с рабочей частотой резонаторов ЛУ ИЯИ и принято $k=0.0856$.\\
Результаты прямого расчета частот в замкнутой КЦ представлены в Таблице 1 и полностью подтверждают сделанные ранее выводы.
\begin{table}[htb]   
\begin{center}
\centering{Table 1: Частоты колебаний полученные прямым решением системы (1). }
\begin{tabular}{|l|c|c|c|c|c|c|c|}
\hline
Mode type        & $N=8$          & все ПРС        & $N=9$       & все ПРС     &  $N=9$         & один ИРС       \\
\hline        
 $q$             & $f_q^{(1)}$    & $f_q^{(2)}$    & $f_q^{(1)}$ & $f_q^{(2)}$ & $f_q^{(1)}$    & $f_q^{(2)}$    \\
\hline
0                & 951.123        &     no         & 951.123     &  no         & 951.693        & 951.693        \\
1                & 953.991        & 953.991        & 953.393     & 953.393     & 956.188        & 956.188        \\
2                & 962.301        & 962.301        & 960.020     & 960.020     & 964.809        & 964.809        \\
3                & 975.155        & 975.155        & 970.448     & 970.448     & 976.803        & 976.803        \\
4                &\textbf{991.000}&\textbf{991.000}& 983.715     & 983.715     &\textbf{991.000}&\textbf{991.000}\\
5                & 1007.643    & 1007.643          & 998.449     & 998.449     & 1005.835       & 1005.835       \\
6                & 1022.430    & 1022.430          & 1012.916    & 1012.916    & 1019.446       & 1019.446       \\
7                & 1032.680    & 1032.680          & 1025.186    & 1025.186    & 1029.914       & 1029.914       \\
8                & 1036.354    &   no              & 1033.440    & 1033.440    & 1035.617       & 1035.617       \\
9                &             &                   & 1036.354    & no          & no             & no             \\
\hline
\end{tabular}
\label{t_freq}
\end{center}
\end{table} 
Для периодической замкнутой КЦ, с одинаковыми ПРС, колебания как при четном $N=8$ - первые два столбца Таблицы 1, так и при нечетном 
$N=9$ - третий и четвертый столбцы Таблицы 1, дважды вырождены по частоте, $f_q^{(1)}=f_q^{(2)}$, за исключением 
колебаний вида $0, q=0$ и $\pi, q=N$.\\ 
Частоты колебаний точно описываются дисперсионным уравнением (3). При нечетном 
$N=9$ колебание вида $\theta =\frac{\pi}{2}$ с частотой $f_{\frac{\pi}{2}} = f_c$ в спектре КЦ отсутсвует.\\ 
При нечетном $N=9$ и введении одного ИРС, пятый и шестой столбцы Таблицы 1, колебание с частотой $f_{\frac{\pi}{2}} = f_c$ в спектре 
КЦ появляется. При этом колебания вида $\pi$ пропадают и значения всех частот смещаются относительно периодического варианта.
Двойное вырождение внутренних видов колебаний $0 < \theta_q < \pi$ присутствует всегда и общее число векторов, 
являющихся решением системы (1) всегда равно $2N$. 
\subsection{Выводы}
В замкнутой периодической КЦ, содержащей $2N$ ячеек, существует $N+1$ вид колебаний.\\
Все колебания внутренних видов $0 < \theta_q < \pi$ дважды вырождены по частоте. \\
Общее число векторов - распределений полей колебаний, существующих в КЦ, равно $2N$.\\
В замкнутой периодической КЦ колебания вида $\theta =\frac{\pi}{2}$ могут существовать только при четном $N$.\\
При нечетном $N$ колебание со знакопеременным распределением по ускоряющим ячейкам на частоте ячеек $f_c$, эквивалентное 
колебанию вида $\theta =\frac{\pi}{2}$, может быть получено введением в КЦ одной инвертирующей ячейки связи.\\
При этом КЦ теряет свойство периодичности и в ее спектре существует $N$ колебаний, дважды вырожденных по частоте.
\section{Практическая реализация КБУС}
Схема практической реализации КБУС, \cite{zav1},\cite{zav2}, отличается от КЦ, описываемой эквивалентной схемой на 
Рис.~\ref{eqv_sc} и показана на Рис.~\ref{mws_models}. Составление эквивалентной схемы для этого устройства  
большого смысла не имеет и это не обязательно для применения прямого метода решения (6). \\
Все УР могут быть одинаковыми, но для формирования КБУС необходимо одновременное использование как ПРС так и ИРС.\\
Пусть УР в устройстве, Рис.~\ref{mws_models}, перенумерованы последовательно, $m=1,2,3,...,N-2,N-1,N$. 
Тогда УР$_1$ связано через ПРС с  УР$_2$ и с УР$_3$ через ИРС. УР$_2$ связано через ПРС с  УР$_1$ и с УР$_4$ через 
ИРС. При $m \geq 3$ УР$_{m}$ связано через ИРС с  УР$_{m-2}$ и с УР$_{m+2}$ тоже через ИРС. 
УР$_{N-2}$ связано через ИРС с  УР $_{N-4}$ и с УР$_N$ через ИРС. УР$_{N-1}$ связано через ИРС с  УР$_{N-3}$ и с УР$_N$ 
через ПРС.\\
Таким образом в устройстве на Рис.~\ref{kbus_pract} всегда содержится четное общее число ячеек $2N$, из них 
$N$ УР, всегда $(N-2)$ ИРС и всегда $2$ ПРС.\\ 
При равенстве частот всех УР, ПРС, ИРС $f_c$ и одинаковом коэффициенте 
связи $k$ спектр, рассчитанные прямым решением задачи (6), содержит $N$ \textbf{дважды вырожденных} колебаний. 
Эквивалентное $\theta =\frac{\pi}{2}$ колебание на частоте $f_c$ существует при любом $N$ и также 
\textbf{дважды вырождено} по частоте.\\
Для обнаружения вырождения колебаний в экспериментах необходимо проводить 
не только измерения распределния поля по УР, что обычно делается при настройке УС, но и измерять распределения полей по РС,
что обычно не делается. 
\section{Прямое численное моделирование частот и распределений полей.}
Применение метода эквивалентных схем было известно и до предложения КБУС, \cite{zav2}. Получившее в дальнейшем 
развитие прямые методы численного моделирования частот и распределений полей, например, MWS CST, \cite{cst},
позволяет непосредственно, без применения аналогий, рассчитать характеристики подобного устройства при разумном $N$. 
\subsection{Настройка ячеек для дальнейших расчетов}
\begin{figure}[htb]
\centering
  \includegraphics*[width=110mm]{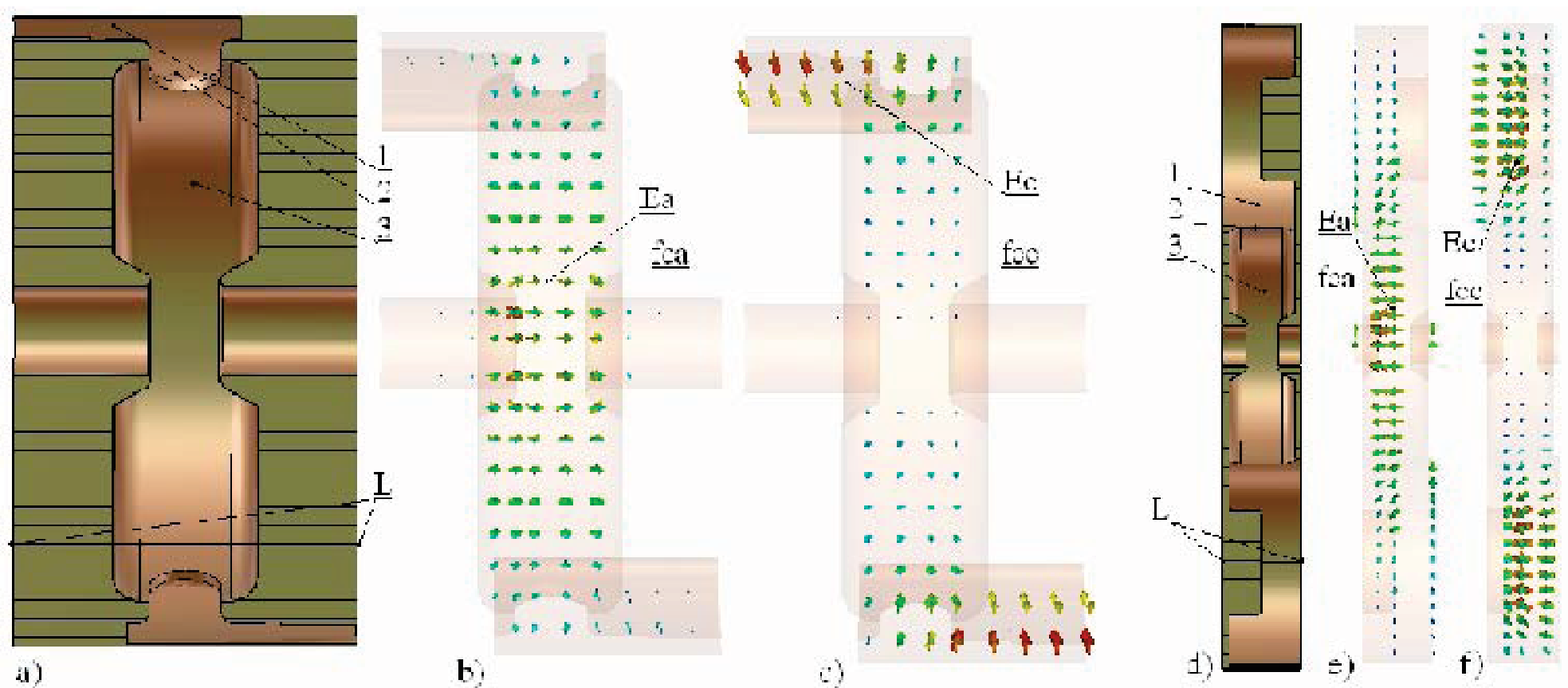}
  \caption{Прямая ячейка связи (a) с распределением поля (b) в сравнении с инвертирующей ячейкой связи (c) с распределением 
поля (d). 1 - ячейка связи, 2 - щель связи, 3 - ускоряющая ячейка.}
	\label{cells_tun}  
\end{figure}

\begin{figure}[htb]
\centering
  \includegraphics*[width=125mm]{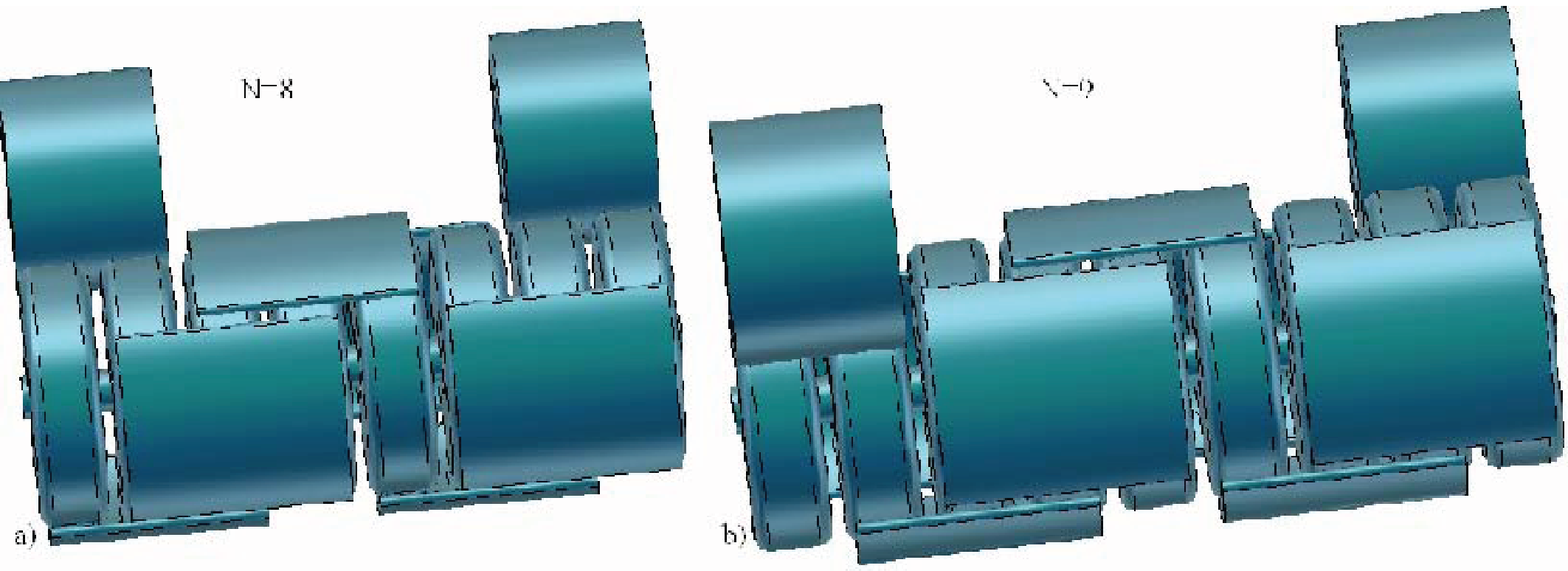}
  \caption{Модели КБУС с $N=8$, (a), и $N=9$, (b) для прямых чиcленных расчетов используя CST MWS.}
	\label{mws_models}  
\end{figure}

Настройка ячеек УР, ПРС и ИРС проведена по стандартной процедуре настройки ячеек и согласования частот, применяемой 
для компенсированных (бипериодических) УС, Рис.~\ref{cells_tun}, с использованием свойств зеркальной симметрии 
настраиваемых ячеек.\\  
При различных типах РС, примененных в КБУС, настройка ячеек как ПРС так и ИРС на заданную частоту достаточно проста в 
численном моделировании. Но при этом трудно обеспечить одинаковые коэффициенты связи для различных типов РС - 
ПРС (боковые ячейки), Рис.~\ref{cells_tun}а, $k_{dc}$  и ИРС (призматические ячейки), Рис.~\ref{cells_tun}, $k_{ic}$. 
Кроме того, различие $k_{dc} \neq k_{ic}$ позволяет рассмотреть более интересный вариант КБУС.\\
Применяемые в дальнейшем ячейки настроены следующим образом. 
Все УР настроены на частоту $f_{ca}≥1.080~MHz$. Ячейки ПРС - боковые - настроены на частоту $f_{ccd} = 991.180~MHz$ 
и обеспечивают коэффициент связи $k_{dc}=0.03293$. Ячейки ИРС - призматические - настроены на частоту $f_{cci} = 991.165~MHz$ 
и обеспечивают коэффициент связи $k_{ic}=0.08561$.\\
 Моделирование полей вырожденных колебаний представляет трудности и для прямых численных методов, поэтому различие в 
частотах УР и РС $\approx 100 kHz$ привнесено \textbf{преднамеренно}. Также преднамеренно обеспесен \textbf{различные} 
величины коэффициентов связи.
Оптимизация ячеек по другим параметрам не проводилась. Цель прямых рассчетов - продемонстирировать дисперсионные свойства 
КБУС.\\
Из настроенных ячеек сформированы две модели с $2N=16$ и $2N=18$ для численного моделирования в CST MWS,
Рис.~\ref{mws_models} и проведены расчеты. Ячейки ПРС смещены на $45~^o$ для уменьшения взаимного влияния щелей связи 
различных ячеек - ПРС и ИРС. 
\subsection{Результаты прямого численного моделирования.}
\begin{figure}[htb]
\centering
  \includegraphics*[width=125mm]{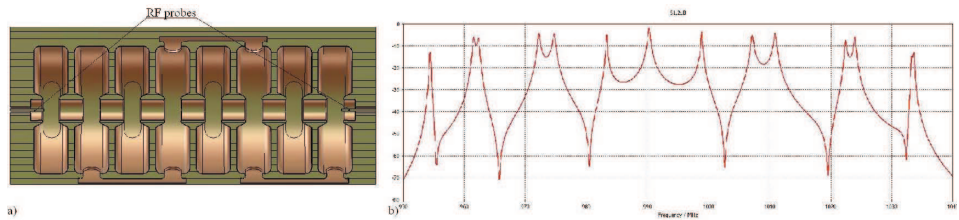}
  \caption{Возбуждение колебаний в резонаторе КБУС, $2N=16$, двумя ВЧ зондами на оси резонатора, (а), и график $S_{12}$ 
в полосе пропускания КБУС, (б).}
	\label{probe_aa}  
\end{figure}

Резонатор КБУС, Рис.~\ref{kbus_pract}, не имеет плоскостей зеркальной симметрии и не обладает свойствами симметрии 
трансляции. Для расчета спектра колебаний приходится рассчитывать резонатор целиком, что требует больших затрат 
ресурсов ЭВМ. Поэтому расчеты проведены с невысокой плотностью сетки тетраэдров на одну ячейку резонатора.\\   
Рассчитанные значения частот в КБУС с $2N=16$ и $2N=18$, полученные как прямым моделированием с CST MWS,
так и методом эквивалентных схем (6), приведены в Таблице 2. Видны как достаточно хорошее совпадение величин, 
полученных независимыми методами, так и проявляющиеся закономерности.\\  
\begin{table}[htb]   
\begin{center}
\centering{Table 2: Частоты колебаний в КБУС, рассчитанные с примененим  CST MWS и методом эквивалентных схем (6), $2N=16$ и $2N=18$.}
\begin{tabular}{|l|c|c|c|c|c|c|c|}
\hline
Mode type         & $N=8$           &  $N=8$         & $N=9$         &  $N=9$            \\
\hline        
 $q$              & $f$, CST MWS    & $f$, Eqv. circ.& $f$, CST MWS & $f$, Eqv. circ.    \\
\hline
1                 & 954.439         & 953.871        & 953.573         & 953.023         \\
2                 & 954.608         & 954.113        & 954.573         & 953.986         \\
3                 & 961.986         & 961.356        & 958.752         & 958.228         \\
4                 & 962.792         & 962.425        & 962.397         & 961.823         \\
5                 & 972.905         & 972.303        & 966.933         & 966.483         \\
6                 & 975.471         & 975.280        & 973.756         & 973.187         \\
7                 & 983.666         & 983.365        & 977.631         & 977.368         \\
8                 &\textbf{991.145} &\textbf{991.089}& 984.128         & 983.841         \\
9                 &\textbf{991.272} &\textbf{991.165}&\textbf{991.155} &\textbf{991.089} \\
10                & 999.307         & 999.075        &\textbf{991.315} &\textbf{991.165} \\
11                & 1007.842        & 1007.773       & 998.786         & 998.577         \\
12                & 1011.637        & 1011.089       & 1005.623        & 1005.484        \\
13                & 1022.776        & 1022.561       & 1010.584        & 1010.098        \\
14                & 1024.446        & 1023.846       & 1018.073        & 1017.757        \\
15                & 1033.238        & 1032.813       & 1023.823        & 1023.284        \\
16                & 1033.675        & 1033.120       & 1028.172        & 1027.665        \\
17                &                 &                & 1033.565        & 1032.974        \\
18                &                 &                & 1034.794        & 1034.201        \\
\hline
\end{tabular}
\label{t_freq_comp}
\end{center}
\end{table} 
Различие в величинах коэффициентов связи ячеек ПРС и ИРС, $k_{dc} \neq k_{ic}$ приводит к снятию вырождения -
расщеплению по частоте для всех видов колебаний, \textbf{кроме рабочего}. Как видно из Таблицы 2, 
колебание 'вида $\theta =\frac{\pi}{2}$' со знакопеременным распределнием поля по ячейкам, остается вырожденным по частоте.
Небольшое различие в частотах этого вида колебаний, см. выделенные величины в Таблице 2, обусловлены преднамеренно 
внесенной расстройкой в частоты УР и РС при настройке ячеек, Рис.~\ref{cells_tun}.\\ 
\begin{figure}[htb]
\centering
  \includegraphics*[width=150mm]{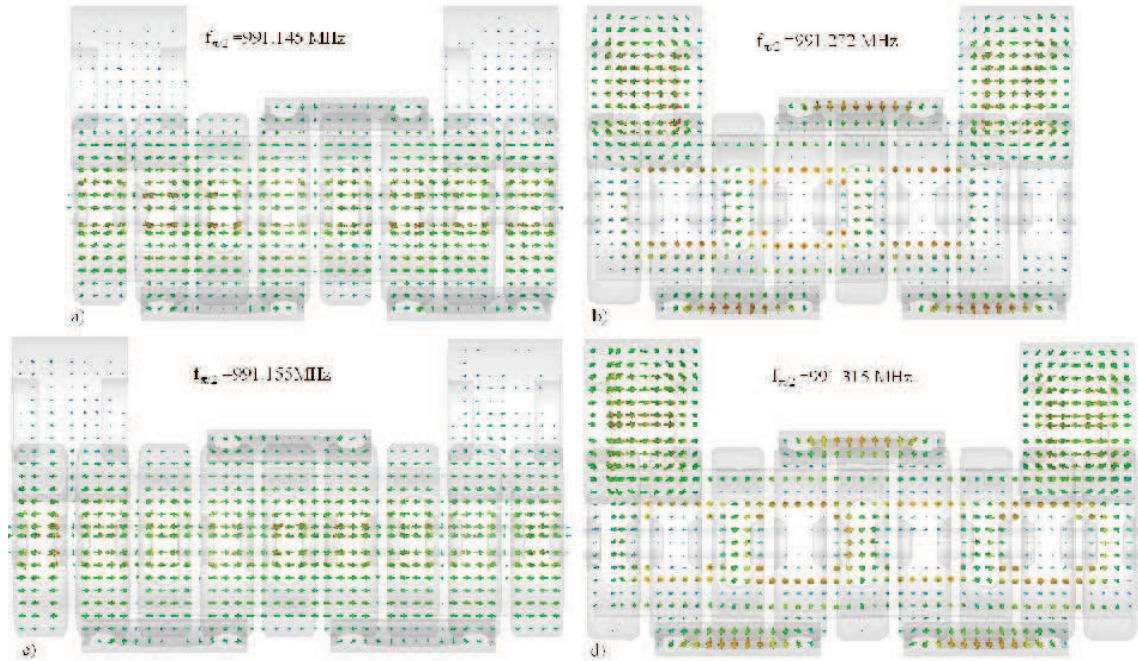}
  \caption{Распределения электрических полей вырожденных колебаний 'вида $\theta =\frac{\pi}{2}$' со 
знакопеременным распределением поля по УР и РС для $2N=16$, (а,б) и $2N=18$, (с,d).}
	\label{89_eac}  
\end{figure} 
На Рис.~\ref{89_eac} показаны распределения электрических полей этих вырожденных колебаний 'вида $\theta =\frac{\pi}{2}$' 
со знакопеременным распределением поля по УР и РС при $2N=16$ и при $2N=18$.\\
На Рис.~\ref{probe_aa} показаны результаты моделирования возбуждения резонатора КБУС, $2N=16$, двумя ВЧ зондами, 
расположенными по оси резонатора, Рис.~\ref{probe_aa}а. На графике $S_{21}$, Рис.~\ref{probe_aa}б, в полосе 
пропускания резонатора ясно видно по крайней мере $14$ (а не $8$) пиков, соответствующих частотам колебаний в резонаторе. 
Но детектировать существование двух вырожденных колебаний 'вида $\theta =\frac{\pi}{2}$' при возбуждении по оси резонатора не 
удается. Для этого по крайней мере один из зондов должен быть в ячейке связи ПРС.\\ 
Вырождение по частоте рабочего колебания является \textbf{необходимым следствием} и всегда присутствует в КБУС при 
настройке УР и РС на одинаковую (рабочую) частоту. Но такая настройка необходима для 
обеспечения симметричного, относительно рабочего, положения по частоте соседних видов колебаний, 
что необходимо для обеспечения стабильности распределния ускоряющего поля, \cite{daw}. Устранить это противоречие, 
находясь в концепции рассмотренной КБУС, невозможно.  
\section{Эффекты при вырожденных колебаниях.}
В линейной цепочке существование колебания вида $\theta =\frac{\pi}{2}$ с полем в РС - колебания связи - 
запрещено граничными условиями на торцах УС. В КБУС этот запрет устранен и колебание связи является полноправным 
колебанием в КБУС на рабочей (или близкой к ней с точностью до погрешностей настройки) частоте.\\ 
\begin{figure}[htb]
\centering
  \includegraphics*[width=125mm]{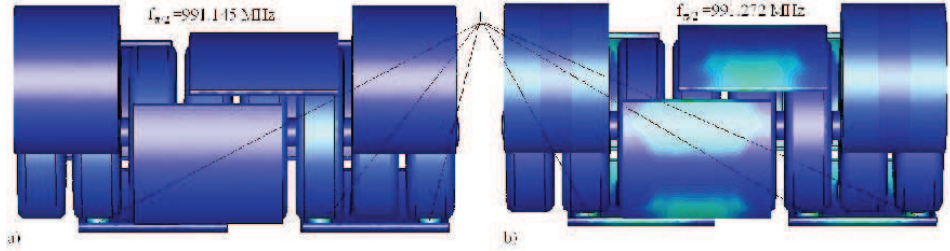}
  \caption{Распределение плотности энергии магнитного поля по поверхности КБУС для вырожденных колебаний 'вида $\theta =\frac{\pi}{2}$'.
1 - участки поверхности с относительно высокой напряженностью магнитного поля.}
	\label{ac_h_coupl}  
\end{figure} 
Эффекты, связанные с вырождением колебаний в резонаторе или наличием в спектре двух близких по частоте колебаний хорошо 
известны в практике СВЧ и рассмотрены в монографиях, см., например, \cite{stein}.\\ 
Математически два колебания, распределения полей которых показано на Рис.~\ref{89_eac} в идеальном резонаторе независимы. 
В реальном резонаторе с отклонениями размеров 
на величину  на погрешности изготовления $\delta V$  всегда происходит связь колебаний, \cite{daw}:
\begin{equation}
\vec E \approx \vec E_n + \sum_{m \neq n} \vec E_m \frac{\omega_m^2 \int_{\delta V}(Z^2_0 \vec H_m \vec H_n^* -\vec E_m \vec E_n^*)dV}
{W_0 (\omega_n^2 -\omega_m^2)},\quad Z_0=\sqrt{\frac{\mu_0}{\epsilon_o}}. 
\label{eq_7}
\end{equation}
В КБУС есть участки поверхности, на которых оба вырожденных колебания, Рис.~\ref{89_eac}, одновременно 
имеют достаточно высокую напряженность магнитного поля. Это участки вблизи щелей связи, Рис.~\ref{ac_h_coupl}. Поэтому 
неизбежные отклонения размеров элементов КБУС при изготовлении неизбежно вызовут связь и смешивание полей этих вырожденных 
колебаний. А при очень близком расположении этих колебаний по частоте, $\frac{f}{\delta f} \approx 10^4$, обеспечивающем 
взаимное перекрытие резонансных кривых колебаний, количественная оценка величины смешивания колебаний не предсказуема.\\
Даже при идеальном изготовлении, $\delta V =0$, связь вырожденных колебаний, осуществляемая на конечной проводимости поверхности 
резонатора, \cite{stein}, будет весьма существенной.\\
Последствия смешивания вырожденных колебаний на ВЧ характеристики КБУС легко прогрозируется. Это отличие получаемых 
изменений частоты ячеек от ожидаемых при настройке КБУС и снижение шунтового сопротивления и добротности рабочего колебания.\\
\subsection{Возможные эффекты влияния на пучок ускоряемых частиц.} 
\begin{figure}[htb]
\centering
  \includegraphics*[width=140mm]{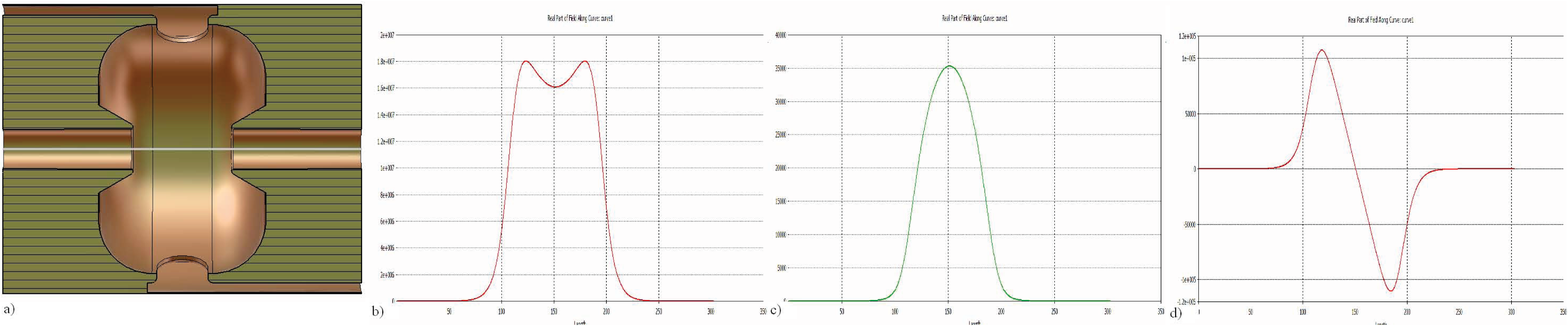}
  \caption{Ячейка КБУС для $\beta=1$, (a), распределение ускоряющего поля $E_z$, (b), распределние поперечной составляющей 
$E_y$ поля рабочего колебания, (с), и распределение продольной составляющей $E_z$ поля ячеек связи на оси КБУС, (d).}
	\label{b1_par}  
\end{figure}

Менее очевидны возможные эффекты влияния КБУС на характеристики пучка ускоряемых частиц. Эти эффекты должны ярче проявляться 
при $\beta \approx 1$ - при применении КБУС в ускорителях электронов, для которых система изначально и разрабатывалась, 
\cite{zav3}.\\ 
На Рис.~\ref{b1_par} показаны ячейка КБУС для $\beta=1$, распределение ускоряющего поля $E_z$, распределние поперечной 
составляющей $E_y$ поля рабочего колебания и распределение продольной составляющей $E_z$ поля ячеек связи на оси КБУС.\\ 
Противоположное размещение щелей связи в УР КБУС приводит с появлению на оси поперечной составляющей электрического 
поля, Рис.~\ref{b1_par}с. Аналогичный эффект известен и для структуры SCS, \cite{scs}. Но в SCS частица при пролете соседних 
зазоров, разделенных на $\frac{\beta \lambda}{2}$, видит противоположно направленную компоненту $E_y$. Частица совершает 
дополнительные поперечные колебания с периодом $\beta \lambda$, которые при пролете длинной секции SCS усредняются.\\
В КБУС ячейки с одинаковым направлением поперечного поля находятся на расстоянии $\beta \lambda$. Происходит синхронное 
взаимодейтсвие частицы с поперечным полем и частица должна получать при пролете КБУС дополнительный поперечный импульс 
в направлении под углом $45~^o$ к оси структуры. Этот эффект носит чисто геометрический характер и не связан с 
вырождением рабочего колебания КБУС.\\
Распределние продольной составляющей $E_z$ поля независимо существующего второго колебания (в ячейках связи) на оси КБУС
показано на Рис.~\ref{b1_par}d. Это колебание обладает меньшей, но не нулевой, величиной эффективного шунтового 
сопротивления. При пролете 
сгустка с несимметирчным распределением частиц в сгустке по фазе или смещении фазы пролета сгустка в целом (отличное 
от нуля значение синхронной фазы) с необходимостью будет происходить взаимодействие, возбуждение, этого независимого 
колебания с пучком.\\ 
Как минимум, возбуждение колебания в РС КБУС будет приводить к отбору энергии от частиц сгустков в конце 
импульса тока пучка и увеличчению разброса частиц по енергии. Как максимум, при длинном импульсе тока частиц 
экпоненциально нарастающая амплитуда колебания в РС КБУС может привести к обрыву импульса тока пучка.  
\section{Заключение}
Идея КБУС основана на концепции, с необходимостью подразумевающей наличие в спектре дважды вырожденных по частоте 
колебаний. Свойства замкнутой кольцевой цепочки резонаторов детально исследованы для кольцевых резонаторов магнетронов 
и существование вырожденных колебаний в ней известно.\\
В практической реализации КБУС представляет собой более сложную, но замкнутую на себя систему связанных ячеек. Для 
построения резонатора КБУС необходимо применение ячеек связи двух типов - двух простых, с одинаковым знаком 
коэффициента связи на различных концах ячейки и $N-2$ инвертирующих, с противоположными знаками коэффициента связи на 
различных концах ячейки. При равенстве частот всех ячеек и равенстве величин всех коэффициентов связи в спектре 
резонатора КБУС содержится $N$ дважды вырожденных по частоте колебаний, где $N$ - число ускоряющих ячеек. Таким 
образом полное число векторов-распределний поля в резонаторе равно $2N$ и равно общему числу ячеек $2N$ в резонаторе.\\
При различных величинах коэффициентов связи прямых и инвертирующих ячеек связи снимается вырождение по частоте всех 
колебаний, за исключением рабочего, со знакопеременным распределением ускоряющего поля в соседних ускоряющих ячейках. 
Сохранение вырождения рабочего колебания с необходимостью следует из настройки всех ячеек, как ускоряющих, так и связи,
на рабочую частоту, что необходимо для обеспечения стабильности распределения ускоряющего поля.\\
В линейной цепочке связанных ячеек возбуждение колебания 'вида $\theta =\frac{\pi}{2}$' в ячейках связи запрещено 
граничными условиями на торцах резонатора. В КБУС этот запрет устранен и колебание 'вида $\theta =\frac{\pi}{2}$' в ячейках связи 
являетя полноправным колебанием резонатора КБУС, существующем также на рабочей частоте.\\
Для обнаружения колебания 'вида $\theta =\frac{\pi}{2}$' в ячейках связи в экспериментах необходимо размещение 
хотя бы одного ВЧ зонда в ячейке связи.\\
Указанные выводы подтверждены как применением традиционного метода эквивалентных схем к резонатору КБУС, так и 
результатами прямого численного моделирования частот и распределний полей в резонаторе КБУС с настроенными ячейками 
с применением современного программного обеспечения.\\
Негативные последствия вырождения колебаний хорошо известны и в практике действующих ускоряющих структур 
вырождение рабочего колебания не допускается.\\ 
Автор выражает  признательность лаборатории DEZY за предоставленную в рамках коллаборации возможность проведения 
численных расчетов с применением CST MWS.


\begin{thebibliography}{99}
\addcontentsline{toc}{section}{\refname}
\bibitem{struct} I.V.~Rybakov et. al., Proposal of the Accelrating Structure for the First Cavity of the Main Part 
of INR Linac,  Proc. RuPAC 2016, p.216, 2017
\bibitem{scs} E.A.~Knapp et. al., Standing wave high energy accelerating structures, 
Rev. Svi. Instr., v. 3, n, p. 979, 1968 
\bibitem{acs} Accelerator Technical Design Report for J-PARC, KEK Report 2002-13, Japan, 2003
\bibitem{daw} 'Accelerating system', in Ion Linear Accelerators, B. Murin, Ed. Moscow, USSR,
Atomizdat, vol. 2, 1978
\bibitem{cds} V.~Paramonov et. al., The PITZ CDS Booster Cavity RF Tuning and Start of Conditioning. 
Proc. 2010 Linac Conference, p. 241, 2011          
\bibitem{nano} ООО Нано Инвест. Ускоряющий резонатор для ММФ. Техническое предложение, 2017. (Prived communication).
\bibitem{knapp_1} E.A.~Knapp et. al., Coupled Resonators Model for Standing Wave Accelerator Tanks, 
Rev. Svi. Instr., v. 38, n, 11, p. 1583, 1967 
\bibitem{strelkov} S.P.~Strelkov, Introduction to the theory of oscillations. The third edition, 2005
\bibitem{mag} В. Н. Шевчик, ред. Электронные приборы сверхвысоких частот. Учебное пособие. Саратовский университет, 
г. Саратов, 2014\\
D.E.~Nagle,~E.A. Knapp, Steady State Behavior of a Ring or of a Chain a Coupled Circuits. LASL report, LA0171, 1961  
\bibitem{zav1} А.А.~Завадцев, Б.В.~Зверев. Новые ускоряющие системы для ЛУЭ со стоячей волной. Писма в ЖТФ, т. 7, в. 21, 
стр. 1332, 1981
\bibitem{nag} http://www.nag.co.uk, release 17, 1999
\bibitem{zav2} А.А.~Завадцев, Б.В.~Зверев. Ускоряющая система. Авторское свидетельство N 852151, 1981. 
\bibitem{cst} http://www.cst.com
\bibitem{stein} В.Б.~ Штейншлегер, Явления взаимодействия электромагнитных волн в резонаторах, Гос. Оборон. Гиз., Москва, 1955
\bibitem{zav3} А.А.~Завадцев, Б.В.~Зверев. Разработка новых ускоряющих систем для специализированных ЛУЭ со стоячей 
волной. Доклады IV Всесоюзного совещания по применению ускорителей заряженных частиц в народном хозяйстве, Л., с. 46, 1981
\end{thebibliography}
\end{document}